\documentclass[3p]{elsarticle}

\usepackage{amsmath}
\usepackage{url}
\usepackage{amsfonts}
\usepackage{amsbsy}
\usepackage{graphicx}
\usepackage{subfigure}
\usepackage{verbatim}
\usepackage{amssymb}

\newcommand{\x}{\ensuremath{\mathbf{x}}}

\renewcommand{\v}{\ensuremath{\mathbf{v}}}
\newcommand{\B}{\ensuremath{\mathbf{B}}}

\renewcommand{\d}{\ensuremath{\partial}}

\newcommand{\ex}{\ensuremath{\mathbf{e}_{x}}}
\newcommand{\ey}{\ensuremath{\mathbf{e}_{y}}}
\newcommand{\ez}{\ensuremath{\mathbf{e}_{z}}}
\newcommand{\ephi}{\ensuremath{\mathbf{e}_{\phi}}}

\newcommand\dd{\partial}

\begin{document}
\title{Resistive
double-diffusive instability in the dead-zones of protostellar disks}
\author[cam1]{Henrik N. Latter\corref{cor1}}
\ead{henrik.latter@lra.ens.fr}
\author[cam1]{Julius F. Bonart}
\ead{bonart.julius@googlemail.com}
\author[cam1]{Steven A. Balbus}
\ead{steven.balbus@lra.ens.fr}

\cortext[cor1]{Corresponding author}
\address[cam1]{LERMA-LRA, 
 \'{E}cole Normale Sup\'{e}rieure, 24 rue Lhomond, Paris 75005, France.}

\begin{frontmatter}
\begin{abstract}

We outline a novel linear instability that may arise in
the dead-zones of protostellar disks, and possibly the fluid interiors of
planets and protoplanets. In essence it is an axisymmetric buoyancy
instability, but one that would not be present in a purely hydrodynamical gas.
The necessary ingredients for growth include
 a negative radial entropy gradient (of any magnitude), 
 weak magnetic fields, and efficient
 resistive diffusion (in comparison with thermal diffusion). The character of the
instability is local, axisymmetric, and double-diffusive, and it attacks
lengths much shorter than the resistive scale.
Like the axisymmetric convective instability, it draws its
energy
from the negative radial entropy
gradient; but
 by utilising the diffusing magnetic field, it can negate the
 stabilising influence of
 rotation.
 Its nonlinear saturated state, while not transporting appreciable angular
 momentum,
 could drive radial and vertical mixing, which may influence the temperature
 structure of the disk, dust
 dynamics and, potentially, planet formation.

\end{abstract}

\begin{keyword}
accretion, accretion disks --- convection --- instabilities --- MHD ---
protoplanetary disks
\end{keyword}

\end{frontmatter}

\section{Introduction}

In this paper, we consider the dynamical behavior of a gaseous system
characterized by rotation (with a strong positive angular momentum
gradient), a weak magnetic field, and a small negative entropy gradient.
As is well known, these ingredients lead to the magnetorotational
instability (MRI), possibly in its more general convective form (Balbus
\& Hawley 1991, Balbus 1995).  It might appear that when resistivity is
present only the longest wavelengths can remain unstable. If the
resistivity is too large then these wavelengths can be longer than the size of the
system, and global stability results.  (Note that the small negative entropy
gradient is stabilized in the hydrodynamical limit by strong rotation.)

Surprisingly, this is not an accurate description of the dynamical
behavior of this system.  Here we show that the presence of resistivity
activates unstable buoyancy modes that rely on the arbitrarily small adverse
entropy gradient as their free energy source.  Moreover, these unstable
modes lie on the same branch of the dispersion relation as the MRI, to
which they revert at small wavenumbers. What is particularly striking
is that at very large wavenumbers, resistivity never damps these modes.
Ultimately, when thermal diffusion is considered, there is in fact
a critical wavenumber above which damping occurs.   But, at least in
protostellar disks and planetary interiors, this still allows for a very
broad range of unstable large wavenumber modes.  (The presence of both
Ohmic resistivity and a much smaller thermal diffusivity is what gives
this instability its `double diffusive' character.)

In this introductory paper, our interest is in a formal presentation
of the properties of the instability in the simplest possible setting.
Our fiducial model will be a protostellar disk.  Such systems are both
cold and dense; excepting the hot regions near the central star and in the
surface layers, much of the gas is poorly ionised, and, as a consequence,
the electrical conductivity too low to develop classical MRI turbulence (Blaes
and Balbus 1994, Gammie 1996, Igea and Glassgold 1999, Sano et al.~2000,
Fleming and Stone 2003, Ilgner and Nelson 2006, Wardle 2007, Okuzumi 2009,
Turner and Drake 2009). 
 Most protostellar
disk models hence consist of an `active' well-ionized envelope, in which
the MRI could in principle be present, encasing an extensive body of
poorly-ionised gas which remains `dead', at least as far as the MRI is
concerned.  The dead region nevertheless may support other dynamics, such
as spiral density waves excited by the turbulence in the surface layers
(Fleming and Stone 2003), streaming instability caused by settling dust
grains (Youdin and Goodman 2005), and flows driven by torques applied by
large-scale magnetic fields (Turner and Sano 2008). Generally,  however,
dead-zones have been considered magnetically `deactivated'. 
Nevertheless, even in the very resistive heart of the
dead-zone we will see that magnetic fields can trigger local small scale instability.

The analysis will be explored in a local model, the details of
which we present in Section 2. In Section 3,  we investigate
the dynamics of linear axisymmetric Boussinesq disturbances by solving
their fifth-order dispersion relation numerically and by obtaining asymptotic
estimates of the growth rates in the double-diffusive limit. The instability
is subsequently related to the idea of magnetostrophic balance.
 In Section 4 our conclusions are drawn and
future work outlined.

\section{Physical and mathematical outline}

Consider a weakly ionized region of a protostellar disk, a nominal
`dead-zone'. A negative radial entropy gradient is assumed to be
present, but the details of how this gradient is established
and maintained do not concern us directly in this calculation: they may
plausibly be present due to a combination of outwardly decreasing
heating from the central star and an outwardly increasing opacity.
The entropy gradient is presumed to be small relative to the
angular momentum gradient of the disk.   More specifically,
the radial Brunt-V\"ais\"all\"a (BV) frequency
\begin{equation}
N^2_R =  - \frac{1}{ \gamma\rho}{\dd P\over \dd R} {\dd \ln P\rho^{-\gamma}
\over \dd R}
\end{equation}
is assumed to be small compared with the epicyclic frequency
\begin{equation}
\kappa^2 = 4\Omega^2 +{d\Omega^2\over d\ln R}.
\end {equation}
Here, $R$ is the radial coordinate in a standard cylindrical system
$(R,\phi, z)$, $\rho$ is the mass density, $P$ is the gas pressure,
$\Omega(R)$ is the angular velocity, and $\gamma$ is the adiabatic
index (ratio of constant pressure specific heat to constant volume
specific heat).  Recall that $\kappa=\Omega$ in a Keplerian disk,
so the magnitude of $N_R$
is assumed much less than an orbital frequency ($N_R$ itself is, of course,
 imaginary here).

We consider the stability of axisymmetric disturbances.  If we
were to treat this highly resistive system as purely hydrodynamic,
we would conclude that it is stable: there is no instability if
$N^2_R+\kappa^2>0$ (the Hoiland-Solberg criterion),
 which for small $|N_R|<\Omega$ is certainly true.
On the other hand, MHD processes are generally also stabilized when the
resistivity is high.  The classical MRI, for example, is banished to
unfeasibly long wavelengths in the presence of strong Ohmic dissipation. 
Consequently, as noted in the Introduction, one might be tempted to conclude
that adding a very small negative entropy gradient to a magnetised resistive disk
 would make very
little difference to the axisymmetric stability.  In fact, it makes a
dramatic difference.

Consider two fluid elements at different vertical locations embedded
in a negative radial entropy gradient.  A weak magnetic field tethers
these two elements to one another.  When these elements are radially
displaced, they will try to bring the magnetic field with them,  but the
field diffuses because of resistivity (which would quell the classical
MRI).  Nevertheless, because of the transient magnetic torque, some finite amount of angular momentum will have
been exchanged between the two elements in the process.
  As a consequence, each element will settle
into a new orbit at a new radial location --- and, most critically,
in a \emph{new entropy environment}.  If thermal diffusion is too
slow to equilibrate the temperature of the displaced fluid with its
surroundings, the elements will continue to move radially, because
they are now buoyantly unstable. Subsequently, the diffusing field will ensure
that these elements will
magnetically connect 
to new fluid parcels, and further angular momentum will be exchanged.
 Thus the cycle continues and an instability is at hand: an
instability that depends upon the transfer of angular momentum between magnetized
fluid elements, yet whose seat of free energy is not the shear, but
rather
the unstable thermal stratification.  Note that this instability is
double-diffusive, relying on the speedy diffusion of angular momentum
(accomplished by diffusing magnetic field) and the slow (or negligible) diffusion
of heat.  This field diffusion counters the stabilising tendency of an
angular momentum gradient by breaking the constraint of specific angular
momentum conservation.

Though buoyantly driven, the instability clearly has features in
common with the MRI. On sufficiently small vertical wavenumbers
(sufficiently spaced fluid elements) the influence of resistive diffusion will be
minor and we recover the MRI.  In fact, we will show that
the double-diffusive instability and the MRI are, in effect, two sides of {\em the
same} instability.  In the presence of a negative radial entropy gradient, one
of the modes that emerges is, at small
wavenumbers, the classical MRI and, at large wavenumbers, the double
diffusive instability.  The instability is suppressed only at extremely
large wavenumbers, when thermal conduction becomes important.
Thus the MRI and the resistive double-diffusive 
instability are intimately linked.

In the following sections this idea is explored using a number of approaches,
mainly within the framework of the linearised Boussinesq equations of resistive
MHD. (An important omission, which we will examine in a later study, is
the electromotive force associated with the Hall term in the induction
equation.)  The local modes under consideration satisfy a 
fifth-order dispersion relation, which we will study in some detail.
It is possible to obtain a number of clean
numerical and analytical results from which further physical insights
emerge.

\subsection{Model equations}

Our model employs the resistive MHD equations, which comprise the
continuity, momentum, and entropy equations, and the
magnetic induction equation with Ohmic diffusion.  As noted,
the Hall effect is also of practical importance, but it is omitted in this
first analysis.  
Ohmic resistivity will exceed viscosity and thermal
diffusion by several orders of magnitude, hence the viscous stress is
dropped.  At least initially, however, it is helpful
to retain the thermal diffusion term, which we model as the divergence of a
radiative energy flux.
The equations are
\begin{align}
\frac{\d\rho}{\d t} + \v\cdot\nabla\rho &= -\rho\nabla\cdot\v, \\
\frac{\d\v}{\d t}+\v\cdot\nabla\v   & = -\frac{1}{\rho}\nabla\left(P + \frac{B^2}{8\pi}\right) -
\nabla\Phi + \frac{(\B\cdot\nabla)\B}{4\pi\rho} \\
\frac{\d\B}{\d t} +\v\cdot\nabla\B &= \B\cdot\nabla\v - \B\nabla\cdot\v+\eta\,\nabla^2\B,
\\
E\,\left(\frac{\d \sigma}{\d t} +\v\cdot\nabla \sigma\right) &= - \nabla\cdot \mathbf{F} + \eta|\nabla\times\B|^2.\label{mark}
\end{align}
Here $\v$ is the
velocity, $E$ the internal energy density, $\Phi$ the gravitational potential,
$\B$ the magnetic field, and $\sigma=\text{ln}P\rho^{-\gamma}$
is the entropy function. The resistivity
is represented by $\eta$, and the radiative energy flux density by $\mathbf{F}$ which is
defined through
\begin{equation}
\mathbf{F}= -\frac{16\,\sigma_B\,T^3}{3\,\rho\,\kappa_\text{op}}\,\nabla T,
\end{equation}
where $T$ is temperature, $\kappa_\text{op}$ is opacity, and $\sigma_B$ is the
Stefan-Boltzmann constant (the latter two
 are not to be confused with the epicyclic frequency $\kappa$ and the entropy
 function $\sigma$). Finally, the equation of state is of an ideal gas
which gives
\begin{equation}
E= \frac{1}{\gamma-1}\,P.
\end{equation}
  In this study, we restrict the analysis
to current-free initial equilibria, i.e.\ $\nabla\times\B=0$, and the last term of equation
(\ref{mark}) will not be used.

\subsection{Linearised equations}

It is assumed that the disk is in near Keplerian equilibrium, with a small
pressure gradient supplementing the star's gravity in the radial force
balance. The equilibrium is $\v_0= R\,\Omega(R)\,\ephi$, $\rho=\rho(R)$, $P=
P(R)$ and $S=S(R)$. It
is also assumed from the outset that the radial gradients of $P$ and
$S$
 are negative and relatively small.
 In addition, weak
 magnetic fields thread the disk, but they are too small to influence
 the equilibrium balances. For simplicity, the magnetic
 field takes a uniform vertical configuration: $\B=B_0\,\ez$. 

In order to examine local disturbances, we adopt the shearing sheet model 
(Goldreich and Lynden-Bell 1965), which represents a small block of disk as a
Cartesian box with $(x, y, z)$ denoting the radial, azimuthal, and vertical coordinates. 
Inside the box $\rho$,
$dP/dR$, and $dS/dR$ are assumed to be constant.

This equilibrium is perturbed by a
planar axisymmetric
Boussinesq disturbance, $\rho'$, $v'_x$, $v'_y$, $B'_x$, $B_y'$, proportional to
$e^{i k z + s t}$, the linearised
equations of which read
\begin{align}
& s\,\v'  = 2\Omega_0\,
u_y'\,\ex -\frac{\kappa^2}{2\Omega_0}\, u_x'\,\ey 
 +\frac{\rho'}{\rho_0^2}\left(\frac{d P}{d R}\right)_0\ex 
 +\frac{i\,k\,B_0}{4\pi\rho_0}\,\B', \label{linu}
\\
& s\,\B' = i\,k\,B_0\,\v' +\left(\frac{d\Omega}{d\ln
      R}\right)_0 B_x'\,\ey -\eta\,k^2\,\B', \label{linB}\\
& s\,\rho' = \frac{\rho_0}{\gamma}\left(\frac{d S}{d
  R}\right)_0 v_x' -\xi k^2 \rho' , \label{linS}
\end{align}
where the subscript $0$ indicates evaluation at the point on which the
shearing sheet is anchored. Hereafter, the subscript 0 will be dropped.
The thermal diffusivity $\xi$ has been introduced which is defined by
\begin{equation}
\xi = \frac{16\,\sigma (\gamma-1)\,T^4}{3\,\gamma\,\kappa_\text{op}\,\rho\,P}.
\end{equation}
 Notice that in the entropy equation \eqref{linS},
 we have used the relation 
\begin{equation}\label{Tpert}
T'= -(T/\rho)\rho',
\end{equation}
 which comes from the perturbed ideal gas law. 
The pressure
perturbation is identically zero for our planar disturbances.
But, in any case, Boussinesq perturbations are in near pressure equilibrium with their
surroundings, and so the fractional pressure perturbation is negligible in
the ideal gas law.

\section{Analysis}

Equations \eqref{linu}--\eqref{linS} present an algebraic
eigenvalue problem for $s$, which yields the following
dispersion relation
\begin{equation} \label{axisdispdim}
s^5 + a_4\, s^4 + a_3\, s^3 + a_2\,s^2 + a_1\,s + a_0 =0,
\end{equation}
with the coefficients
\begin{align}
a_4 &=  2\eta\,k^2 + \xi\,k^2, \\
a_3 &=  N^2_R+\kappa^2+ 2 v_A^2\,k^2 + \eta^2\,k^4 + 2\eta\xi\,k^4, \\
a_2 &=  2N^2_R\,\eta\,k^2 + 2(\eta + \xi)v_A^2\,k^4
+(2\eta+\xi)k^2\,\kappa^2 +\eta^2\xi\,k^6, \\
a_1 &=  v_A^2\,k^2\left(\frac{d\Omega^2}{d\ln R}\right) + v_A^4\,k^4 +
(\eta^2\,k^2 + v_A^2)\,k^2\,N^2_R 
  +(2\eta\xi + \eta^2)\,k^4\,\kappa^2 +  2\eta\xi v_A^2\,k^6, \\
a_0 &= N^2_R\,\eta\,v_A^2\,k^4+ \xi\,v_A^4\,k^6  + \eta^2\xi\,k^6\,\kappa^2 +
\xi\,v_A^2\,k^4\left( \frac{d\Omega^2}{d\ln R}\right), \label{a0}
\end{align}
where the Alfv\'en speed is defined through $v_A= B_0/\sqrt{4\pi\rho}$.
Equation \eqref{axisdispdim} is a
special case of the dispersion relation calculated by Menou et al.~(2004), which also accounts
for viscosity, vertical variation in $\Omega$, $P$, and
$S$, and radial wavenumbers. On the other hand, by taking $\eta=\xi=0$ we recover the
ideal MHD relation of Balbus (1995), which yields the two convective
modes and the two
 convective-MRI modes. The dispersion relation in
this case is fourth order, which means that the inclusion of magnetic diffusion (and
thermal diffusion) leads to the
emergence of a new fifth mode.

From simply inspecting the sign of the last coefficient $a_0$ it is 
already clear that the system possesses
novel stability behaviour when $\xi$ is small.
If we consider scales upon which the thermal diffusion has little
influence, the stability discriminant is the first term on the right side of \eqref{a0}.
It follows that stability is linked to the sign of $N^2_R$, which means that the
Schwarzchild criterion is recovered --- \emph{the stabilising effect of rotation has
vanished}. Notice also that the discriminant combines the three elements
essential to the new instability:
 resistivity, entropy stratification, and magnetic fields. If any one
of these is missing $a_0$ vanishes and the systems' stability properties
revert to the familiar situation of Balbus (1995), which is governed by
the coefficient $a_1$.

In the following subsections we adopt
 a number of approaches in order to describe the mathematics and physics of this
 double-diffusive mode more fully. First, we briefly discuss the main parameters that
 appear in the model and estimate realistic values they may take in
 protostellar disks. Next we solve the dispersion relation
 Eq.~\eqref{axisdispdim} numerically, and obtain an analytical
 asymptotic result in the regime of interest. Following that, we show how the
 mode relies on a generalisation of `magnetostrophic balance' in order to
 function. Finally, we give a more physical explanation of the
 mechanism of instability.

\subsection{Fiducial parameters and lengthscales}
 Time and wavenumber units are chosen so that $\Omega=1$ and
 the Alfv\'en wavenumber, defined through $k_A = \Omega/v_A$, is also $1$. 
The Alfv\'en length is denoted by $l_A$ and
 is defined by $l_A=2\pi/k_A$. Consequently, the
 governing parameters of the dispersion relation comprise the scaled squared BV
frequency $N^2_R/\Omega^2$, the scaled squared epicyclic frequency $\kappa^2/\Omega^2$,
 the Roberts number $q$, and the Elsasser number $\Lambda$. The latter two
 are defined by
\begin{equation} \label{RobandElsie}
q = \frac{\xi}{\eta}, \qquad \Lambda = \frac{v_A^2}{\eta\,\Omega}.
\end{equation}
In order to estimate $q$ we use the following formulas for the
resistivity and thermal diffusivity
\begin{align}\label{eta}
\eta &= 2.34\times
10^{17}\,\left(\frac{T}{100\,\text{K}}\right)^{1/2}\left(\frac{10^{-14}}{x_e}\right)\,\text{cm}^2\,\text{s}^{-1},
\\
\xi &= 2.39\times
10^{12}\,\left(\frac{\rho}{10^{-9}\,\text{g}\,\text{cm}^{-3}}\right)^{-2}\left(\frac{T}{100\,\text{K}}
\right)^3 \,\left(\frac{\kappa_\text{op}}{\text{cm}^2\,\text{g}^{-1}}  \right)^{-1}\,\text{cm}^2\,\text{s}^{-1},
\end{align}
where $x_e$ denotes electron fraction (Blaes and Balbus 1994). In the latter
expression we
have used $\gamma=7/5$, and endowed the gas with a mean
molecular weight of 2.3. According to the commonly used minimum mass model, at a few
AU the disk may be characterised by $T\sim 100$K and $\rho\sim
10^{-9}\,\text{g}\,\text{cm}^{-3}$ (Hayashi et al.~1985), while at these low temperatures
$\kappa_\text{op}\sim 1\,\text{cm}^2\text{g}^{-1}$ (Henning and Stognienko
1996). The ionisation fraction $x_e$, on the other hand, is poorly constrained
and extremely sensitive to dust grain size and abundance, the sources of
ionisation, 
and vertical location. Various model show it ranging over many orders of
magnitude: from
$10^{-13}$ to values as low as $10^{-19}$ (Sano et al.~2000, Ilgner and
Nelson 2006, Wardle 2007, Turner and Drake 2009). Given this uncertainty, we
estimate the
 Roberts number by $q\sim x_e/10^{-9}$. This suggests an upper limit for $q$ of
 $10^{-4}$ and an extreme lower limit of $10^{-10}$.

The Elsasser number requires knowledge of the strength of the
 latent magnetic field, which we parameterise by the plasma beta:
\begin{equation}
 \beta=\frac{
 2\,c^2}{v_A^2},
\end{equation}
 where $c$ is the gas sound speed. This gives
\begin{equation}\label{lamm}
 \Lambda \sim \frac{1}{\beta}\,\left(\frac{H^2\Omega}{\eta}\right) 
\end{equation}
with $H= c/\Omega$ the disk scale height. The bracketed term in Eq.~\eqref{lamm} is
the magnetic Reynolds number, for which we can obtain a rough estimate using equation
Eq.~\eqref{eta} and assuming $H=0.1R$ and $R\approx 5$ AU. For $T\sim 100\,$K, this
presents $\Lambda\sim \beta^{-1}(x_e/10^{-15})$. So one may expect $\Lambda$ to
vary between $10^{-1}$ and $10^{-6}$ or even lower.

Finally, it is necessary to estimate the squared BV frequency $N^2_R$,
which represents the entropy gradient from which the double-diffusive 
modes derive their energy. Most $\alpha$-disk models in fact yield a positive BV
frequency (Balbus and Hawley 1998), but this is because they omit the central
star's radiation field which will tend to drive a negative radial temperature
gradient. However, it is difficult to quantify how this central heating is
mediated by the turbulent inner radii of the disk, and the dead-zone itself,
and so it is unclear which estimates one should use for the magnitude of $N^2_R$.
 We simply assume that $N^2_R/\Omega^2$ takes values between $-0.1$
and $-0.01$.
Because of the pressure gradient the background flow $\Omega$ is not strictly Keplerian, however
the deviation should be minimal and so we set $\kappa\approx\Omega$. 

 The Alfv\'en length scales as
$$l_A\sim \beta^{-1/2}H.$$
 Therefore modes that possess a wavelength
comparable or smaller than the Alfv\'en length may fit comfortably into the
disk when the fields are sufficiently sub-equipartition. In contrast the
resistive length, defined by $l_\eta= \eta/v_A$, scales
like 
$$l_\eta\sim \Lambda^{-1}l_A.$$
There can be no classical-MRI on scales less than $l_\eta$. Using the above
 estimate for $\Lambda$ we derive $l_\eta\sim \beta^{1/2}(10^{-15}/x_e)\,H$. 
Then, for realistic parameter values
 $\beta>10^2$ and $x_e<10^{-14}$, the following ordering is established:
\begin{equation} \label{ordering}
\lambda\,\text{(MRI)}\,>\,\, l_\eta\, >\, H\, >\, l_A\,.
\end{equation}
Thus the classical MRI modes will not occur in the main body of the disk in
most cases,
giving rise to a dead-zone. The double-diffusive
instability, on the other hand, will be present on scales from $l_\eta$ down to a critical
length $l_c$ upon which radiative diffusion becomes important. This critical length
can be estimated by setting the constant coefficient $a_0$ in \eqref{a0} to zero. We find
\begin{equation}\label{kc}
l_c\sim q^{1/2}\Lambda^{-1}\,\frac{\Omega}{|N_R|}\,l_A\,\sim
\beta^{1/2}
\left(\frac{\Omega}{|N_R|}\right)\left(\frac{10^{-21}}{x_e}\right)^{1/2}\,H.
\end{equation}
Unless magnetic fields and the entropy gradient are exceptionally weak,
 this scale $l_c$ should be less than the scale height.
 We conclude that unstable double-diffusive modes will generally fit into the dead-zone.

\subsection{Numerical solutions}

In this section, the dispersion relation \eqref{axisdispdim} is solved
numerically. Once the parameters $\Lambda$, $\kappa^2/\Omega^2$,
$N^2_R/\Omega^2$, and $q$ are stipulated, we
may compute the growth rate $s$ as a function of vertical wavenumber $k$. 
Throughout we set $q$ to be a constant, $q=10^{-6}$, and $\kappa/\Omega=1$,
 while
varying $\Lambda$ and $N^2_R/\Omega^2$. Hence, the Elsasser number $\Lambda$ may be
interpreted as a surrogate for magnetic field strength. The results of this subsection are limited to the
 MRI/double-diffusive mode, which is the only mode that grows, and we neglect the other
 four decaying modes.

\begin{figure}
\scalebox{0.7}{\includegraphics{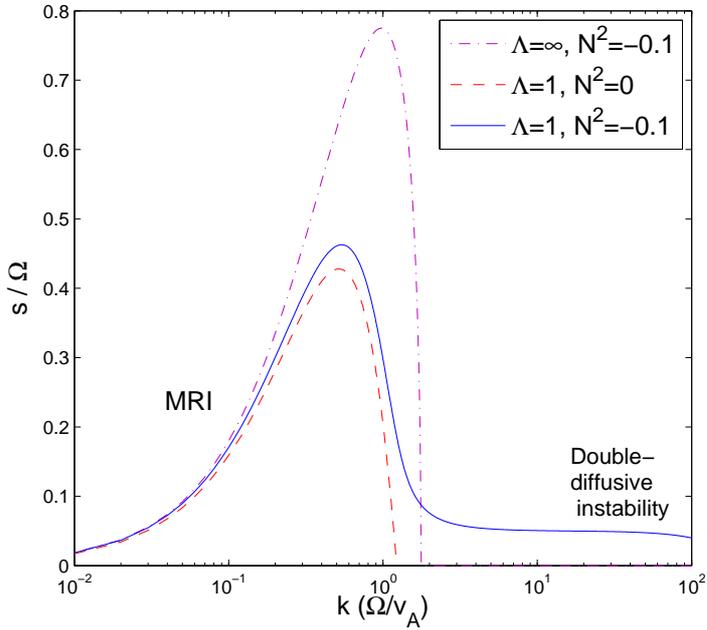}}
\caption{Growth rate $s$ of the unstable MRI/double-diffusive mode as a
  function of $k$ for three illustrative cases. The purple dashed-dotted line represents
  the ideal MHD configuration $\Lambda=\infty$ with $N^2_R/\Omega^2=-0.1$. Here the
  unstable mode is the convective-MRI and it is extinguished when $k/k_A\gtrsim \sqrt{3}$. The
  red dashed line represents resistive MHD in the absence of a radial entropy
  gradient; here $\Lambda=1$ and $N^2_R=0$. Instability is extinguished on
  wavelengths roughly below the resistive scale. The solid blue line represents the
  case with both resistivity and the negative entropy gradient: $\Lambda=1$
  and $N^2_R/\Omega^2=-0.1$. As is plain, instability continues into 
  short scales; this is the double-diffusive instability. In the non-ideal
  case, the ratio of thermal to resistive diffusivity is set to 
  $q=10^{-6}$.}
\end{figure}

Figure 1 illustrates the remarkable stability behaviour unlocked by the
combination of resistivity and a
negative entropy gradient. The purple dotted-dashed curve
represents the MRI in ideal MHD ($\eta=0$ but with $N^2_R\neq 0$) and the red 
dashed curve represents the
MRI in resistive MHD but with no radial stratification ($\eta\neq 0$ and $N^2_R=0$). In both
cases instability is extinguished on relatively long scales: near an Alfv\'en length in
the former case, and near the resistive length in the
latter case. (Because here $\Lambda=1$ the two scales are comparable.) 
However, when resistivity \emph{and} stratification
are present ($\eta\neq 0$ and $N^2_R\neq 0$) instability extends to much
shorter scales, as witnessed by the blue solid curve in Fig.~1. We identify this
`extension' of the MRI as the double-diffusive instability.
 The MRI
transforms smoothly into the new instability upon a transitional scale comparable
to $l_\eta$; both instabilities occur on
the \emph{same branch} of the dispersion relation. When the MRI
mechanism is quenched (because of resistivity),
 the double-diffusive mechanism takes its place. Note
also that the
growth rate of the latter is effectively independent of wavenumber for a broad
range of $k$.

\begin{figure}
\scalebox{0.7}{\includegraphics{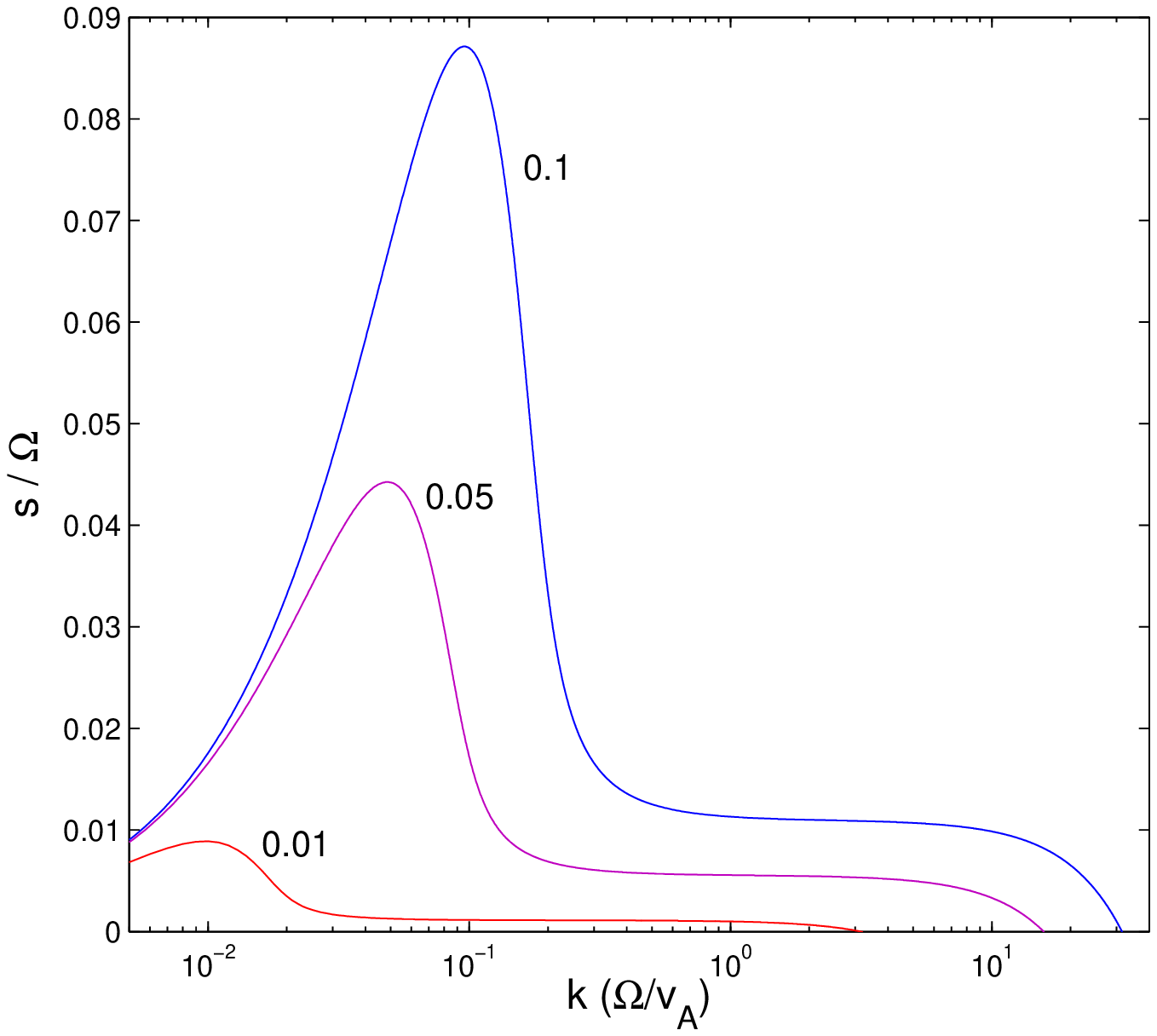}}
\caption{Growth rates of the double-diffusive mode $s$ as a function of $k$
  for different Elsasser number $\Lambda$. The blue line represents the $\Lambda=0.1$ case,
  the purple line, $\Lambda=0.05$, and the red line, $\Lambda=0.01$. The squared
  BV frequency is $N^2_R/\Omega^2=-0.1$ for these cases and the
  Roberts number is $q=10^{-6}$.
 At very long scales, $k\ll k_\eta\sim\Lambda k_A$, we recover the familiar
  MRI mode, but as $k$ approaches $k_\eta$ this mode morphs into the double-diffusive
  mode which persists until $k_c$.}
\end{figure}

\begin{figure}
\scalebox{0.7}{\includegraphics{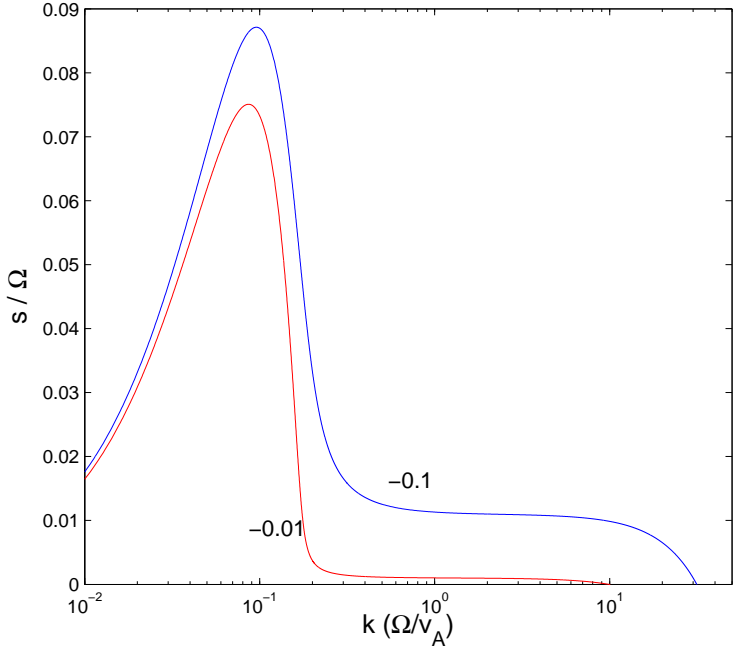}}
\caption{Growth rates of the double-diffusive mode $s$ as a function of $k$
  for different squared BV frequency $N^2_R$. The blue line represents the
 $N^2_R/\Omega^2=-0.1$ case,
 and the red line, $N^2_R/\Omega^2=-0.01$. The Elsasser number
  is $\Lambda=0.1$ and the
  Roberts number is $q=10^{-6}$.}
\end{figure}

In Figs 2 and 3, we present the behaviour of the growth rates as
$\Lambda$ and $N^2_R$ vary. In both case, the smaller $\Lambda$ and
$N^2_R$, the smaller the double-diffusive growth rate.
Numerically, it appears to scale as
$$ \frac{s}{\Omega}\sim \left(\frac{-N^2_R}{\Omega^2}\right)\,\Lambda.$$
We show this analytically in Section 3.2 with an asymptotic analysis. Being the
product of two small quantities, realistic growth rates will be much smaller
than the orbital frequency, though still dynamically significant. For example,
taking $\Lambda=10^{-3}$ and
 $N^2_R/\Omega^2=10^{-2}$ furnishes
 a growth rate of $s\sim 10^{-5}\Omega$,
 and thus an e-folding time of some 15,000 orbits.

The range of unstable wavenumbers associated with the double-diffusive
 instability
 is governed by the
Roberts number $q$ and $N^2_R$. It extends from, roughly, the resistive scale $k_\eta\sim\Lambda$
down to $k_c\sim |N_R/\Omega|\Lambda q^{-1/2}$ (in units of $k_A$).

\subsection{Asymptotic analysis}

In this section we derive an analytic expression for the
 growth rates in the double-diffusive regime:
  when wavelengths are much less than the
 resistive scale $l_\eta$ but much longer than the cutoff scale $l_c$. These
  modes are thus heavily influenced by electrical diffusion but are relatively
 insensitive to radiative diffusion. An asymptotic
 analysis of very long wavelength modes, which connects this paper to the
  local stability results of Balbus (1995), is presented in Appendix A.

In order to calculate modes in the double-diffusive regime, we set
 set their lengthscale to a value of order the Alfv\'en length, 
i.e.\  $k\sim k_A$.
In addition, a small scaling parameter $\epsilon$ is introduced so that $0<\epsilon\ll 1$.
 From the outset we assume that both $\Lambda$ and $N^2_R/\Omega^2$ are
 equally small
and of order $\epsilon$. Subsequently, the growth rate $s$ is
 expanded in powers of $\epsilon$, substituted into the dispersion relation,
 and then terms of equal order are collected.
This procedure permits the calculation of the
 leading order terms of all five modes.

We find that the two hydrodynamic convective modes oscillate at the epicyclic
frequency at leading (zeroth) order and decay at a rate
$-\Lambda\,\Omega$. There exist two other modes, on the other hand, which both decay at the
rate $-\Lambda^{-1}\,(k/k_A)^2\,\Omega$. 

The fifth mode is the double-diffusive mode and it
possesses a growth rate that scales like $\epsilon^2$. It can be determined
 by balancing the last two terms in the dispersion relation
 Eq.~\eqref{axisdispdim}. We find, to leading order,
\begin{equation} \label{dds}
s= -\frac{N^2_R}{\kappa^2}\,\Lambda\,\Omega.
\end{equation}
Stability of the mode is governed by the classical Schwarzchild criterion. So when
  $N^2_R<0$, the mode grows. Note also that the growth rate is independent of the
wavenumber $k$ in this regime. The expression
 \eqref{dds} is, in fact, quite general and holds throughout most of
 the double-diffusive range of $k$, as confirmed by the numerical solution (Figs 2 and
 3). When $k$ becomes small, however, the analysis breaks down and the growth
 rate connects to the classical-MRI curve. And when $k$ is very large, thermal
  diffusion stabilises the mode.

\subsection{Magnetostrophic balances}

In this subsection we characterise the instability in a physically illuminating
way, by showing how it relies on certain steady balances.
 While the resistive MRI (or magnetostrophic MRI) works by balancing the Lorentz and
inertial forces in the momentum equation (as discussed in Petitdemange et
al.~2008), the double-diffusive instability requires \emph{in addition} the balancing
of advection and diffusion in the induction equation.
 Recognising this fact helps simplify some of the algebra
of the previous section, and adds support to the previous asymptotic estimate.

\subsubsection{Balancing the momentum equation}

Consider the linearised momentum equation \eqref{linu}. Assume the
left hand side is subdominant.
 This means that to leading order we have a steady balance between the Lorentz
 and inertial forces (magnetostrophic balance),
 also relevant to the Earth's core (Petitdemange et al.~2008).
 By keeping the growth rate explicit
 in the induction and entropy equation, but setting $\xi=0$,
 we may derive a (reduced) cubic
 dispersion relation,
\begin{align}
&\kappa^2\,s^3 + 2\kappa^2 \eta\,k^2\,s^2  +
 \left[(N^2_R+\widetilde{\Omega}^2)v_A^2\,k^2+ 
(v_A^4+\kappa^2\eta^2)k^4  \right]s  +N^2_R\,\eta\,v_A^2\,k^4 \,=\,0,
\end{align}
with $ \widetilde{\Omega}^2 = (d\Omega^2/d\ln R)$.
While an approximation, this equation
captures the physical essence of the two MRI modes and the double-diffusive mode. 
We may obtain an analytic expression by solving for $k^2$, which, after
inversion, yields the growth rates of these modes.
For purely real $s$:
\begin{align}
k^2_{\pm} &= -\frac{2\kappa^2\eta\, s^2+(N^2_R+\widetilde{\Omega}^2)v_A^2\,s}
{2(v_A^4s+\eta N^2_R
  v_A^2 + \kappa^2\eta^2 s)}
  \pm\frac{v_A\,s\sqrt{4\kappa^2\eta\widetilde{\Omega}^2s
  +(N^2_R+\widetilde{\Omega}^2)^2\,v_A^2 - 4\kappa^2\,v_A^2\,s^2 }}
{2(v_A^4s+\eta N^2_R
  v_A^2 + \kappa^2\eta^2 s)}. \label{cubic}
\end{align}
\begin{figure}
\scalebox{0.7}{\includegraphics{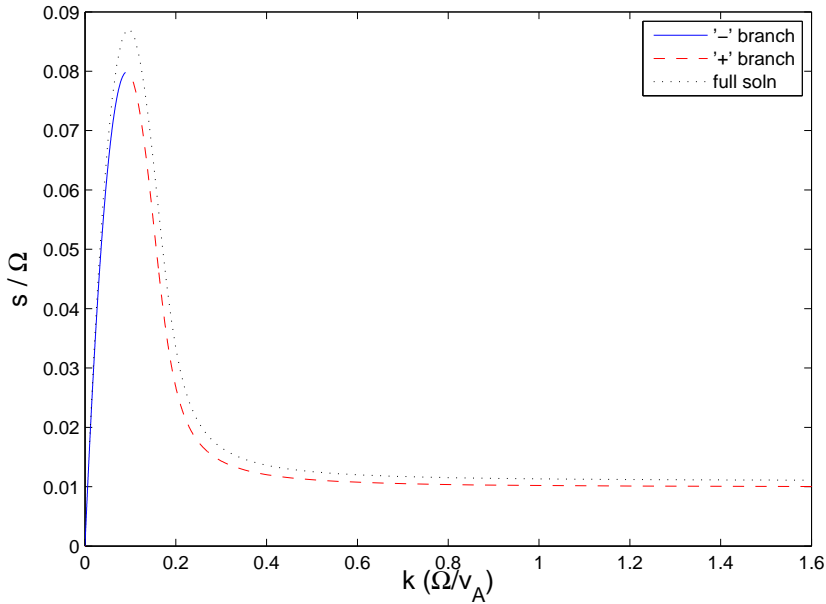}}
\caption{Growth rate of the purely real unstable mode $s$ as a function of $k$
  as determined from the inverse of \eqref{cubic}. The red dashed line indicates
  the plus branch and the solid blue line the minus branch in \eqref{cubic}. The
  dotted black curve represents the full solution to the fifth-order dispersion
  relation \eqref{axisdispdim}.
  The parameters chosen are $N^2_R/\Omega^2=-0.1$, $\Lambda=0.1$,
  $\widetilde{\Omega}^2/\Omega^2=-3$, $\kappa/\Omega=1$.}
\end{figure}
There are thus two branches, which we plot in Fig.~4 
for representative parameters. The solid blue line indicates the `minus' branch and
the dashed red line the `positive' branch. In addition, we plot the full solution to
the fifth-order dispersion relation \eqref{axisdispdim} (the dotted black
line).
 As is clear, the two
branches of the reduced magnetostrophic solution present a reasonable approximation
to the actual MRI and double-diffusive modes for all $k$. This strengthens the
claim that magnetostrophic balance is central to the mechanism of instability in both
cases.

\subsubsection{Balancing the induction equation}

It is also instructive to consider the limit in which
 the time derivatives are negigible in the induction equation
as well. We thus set the left-side of the linearised equation \eqref{linB} to
zero. This means that in addition to the steady balance between the Lorentz
force and the inertial forces in the momentum equation, there is a steady
balance between magnetic diffusion and magnetic advection/distortion. This
leaves only one time-derivative in the linearised equations
 \eqref{linu}--\eqref{linS}.
As a result, the growth rate is
straightforward to calculate:
\begin{equation} \label{rmsbal}
s= \frac{-N^2_R\,\eta\,v_A^2\,k^2}{\widetilde{\Omega}^2 v_A^2 + (v_A^4+\eta^2\kappa^2)\,k^2}.
\end{equation}

This expression yields the (scaled) asymptotic growth rate of the double-diffusive
instability \eqref{dds} in the limit of large resistivity and for 
$k \gtrsim v_A/\Omega$. On the other hand, for sufficiently small $k$ we
obtain the growth rate of a decaying purely resistive mode described in the
Appendix by Eq.~\eqref{resis}. These two modes, however, occur on different solution branches and
the transition between them, in the above, is discontinuous.

The two steady balances upon which the double-diffusive mode exploits, may be understood
 as a cancelling of the forces associated with the magnetic field and differential rotation.
Differential rotation will attempt to stretch out a field line, and magnetic
diffusion will relax the ensuing magnetic tension. In the meantime,
however, what magnetic tension is generated can counterbalance the inertial forces
in the momentum equation. This is important because these inertial forces
 would otherwise stabilise the mode.

\subsubsection{Angular momentum and heat transport}

The solution obtained in the previous subsection presents a convenient
framework to compute the angular momentum and heat fluxes aroused by a single
 linear mode. These may offer an insight into the direction of these fluxes in
 the saturated nonlinear state.

The thermal heat flux associated with a single Fourier mode of the
double-diffusive instability we define as
\begin{equation}
\mathcal{F}_\text{T} = 2\,\text{Re}(u_x'\, \overline{T}') = -2\,\text{Re}(u_x'\,\overline{\rho}'),
\end{equation}
where the overline indicates the complex conjugate. From the linearised
equations, with $s$ negligible in the momentum and induction equations, we obtain
\begin{equation}
\mathcal{F}_\text{T} = 2\, |u_x'|^2\,\left(\frac{-\rho\,\d_R S}{\gamma\,s}\right).
\end{equation}
With $s>0$ and positive, from Eq.~\eqref{rmsbal}, and a negative radial
entropy gradient, the double-diffusive mode always transports heat outwards,
at a rate proportional to the gradient itself. This is as expected: the mode
will act against the destabilising condition from which it arose.

The angular momentum flux we define as
\begin{equation}
\mathcal{F}_\text{h} = 2\left[\text{Re}(u_x'\,\overline{u}_y') 
-\frac{1}{4\pi\,\rho_0}\,\text{Re}(B_x'\,\overline{B}_y') \right].
\end{equation}
Using the eigenfunction of the double-diffusive instability, this can be reexpressed as
\begin{equation}
\mathcal{F}_\text{h} =
-|u_x'|^2\left[\frac{1}{\Lambda}\left(\frac{\kappa^2}{\Omega^2}\right)
- 4\Lambda\,\frac{\Omega^2}{v_A^2\,k^2}\right].
\end{equation}
The double-diffusive mode works on scales $k\gtrsim v_A/\Omega$, and we take
$\Lambda$ to be small. Thus the first term (associated with the Reynolds
stress) in the square brackets will
dominate the second, which ensures that angular momentum is transported
inwards. However, for very small $k$, i.e.\ for the decaying resistive mode,
 the situation is reversed.

\begin{figure}
\scalebox{0.7}{\includegraphics{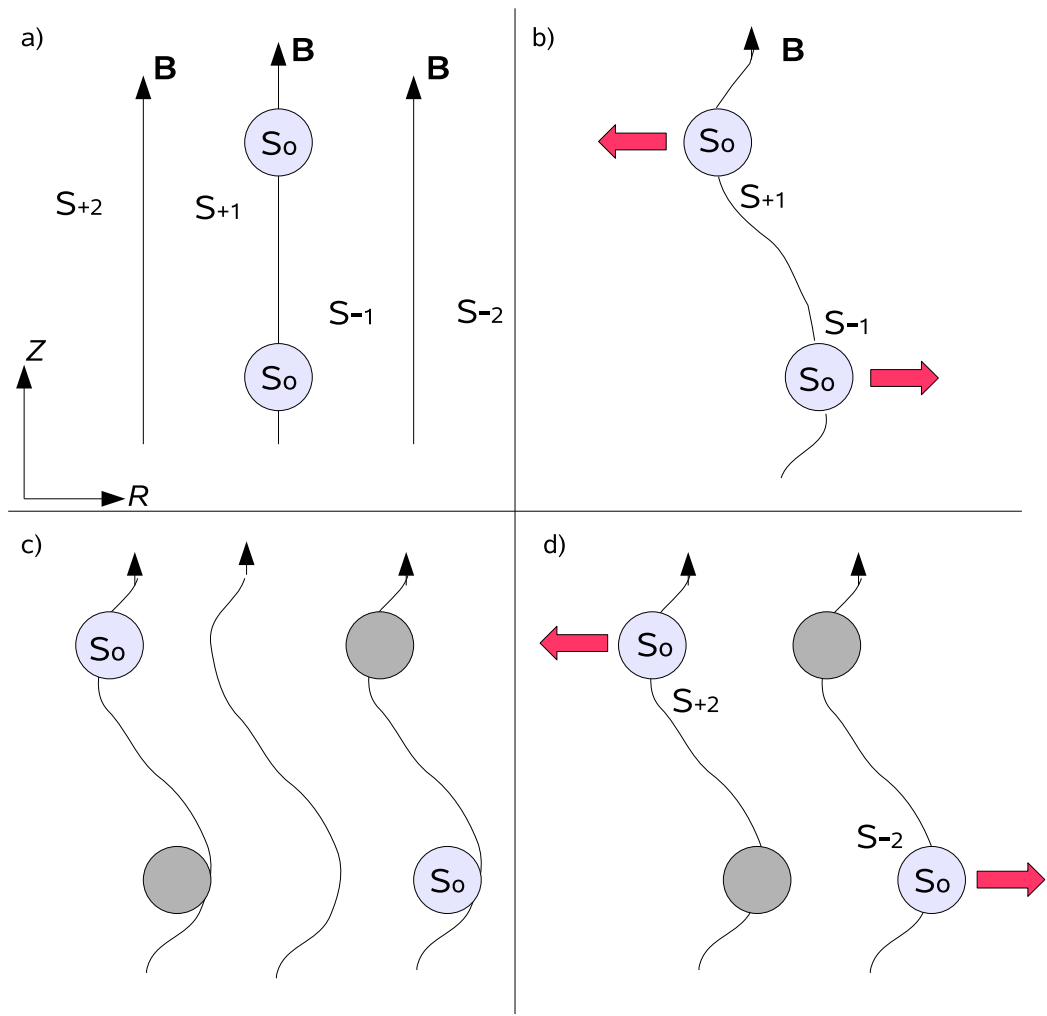}}
\caption{Four panels showing schematically the physical mechanism of the
  double-diffusive instability. Panel (a) presents two fluid blobs, embedded
  in a negative entropy gradient: $S_{+2}>S_{+1}>S_0>S_{-1}>S_{-2}$. The fluid
  elements are separated vertically but connected by a
  magnetic tether. The following panels, in order, show the development of a
  horizontal perturbation in time, under the influence of speedy angular
  momentum transport (by the diffusing magnetic field) and the radial entropy
  gradient. See Section 3.5}
\end{figure}

\subsection{Physical description of the unstable mode}

 Figure 5 presents a schematic diagram of how the double-diffusive instability
  works in a physically intuitive way, developing the ideas put forward at the
  beginning of Section 2.
Panel (a) shows two fluid blobs at
  different vertical locations but at the same radial location (the violet circles). They are 
  connected by a vertical magnetic field
  line. The two blobs are embedded in a negative radial entropy gradient, but
  they
 possess the same entropy $S_0$ as each other, because they inhabit the same
  radius. Fluid at smaller radii (to
  their left) possesses greater entropy, $S_{+1}$ and $S_{+2}$, and fluid at greater radii (to their right)
  possesses lesser entropy, $S_{-1}$ and $S_{-2}$. In panel (b) the blobs are perturbed
  horizontally, initially deforming their magnetic tether. Because of the large
  resistivity, the blobs will slip through the magnetic field, and so the
  magnetorotational instability does not set in.
  Nevertheless, the field will permit some amount of 
angular momentum to be exchanged between the two blobs
  and they adopt new orbits at their new radii. But
  the displaced fluid blobs now possess a
  different entropy to their surroundings: too little entropy in
  the case of the upper blob ($S_0<S_{+1}$), and too much in the case of the
  lower ($S_0>S_{-1}$).
Unless thermal diffusion acts
  rapidly to equilibriate them with their ambient gas, they cannot sustain their current
  positions. Instead
 they will continue moving radially (pink arrows).
  This motion will bring them into contact with new diffusing magnetic field lines,
  panel (c),
  which will tether them to new fluid partners (the grey circles). 
  Now the new magnetic field lines will facilitate another transient torque, 
which will transfer more angular momentum. As earlier, the violet blobs adopt
  new radii but the disparity between their entropy and their environments'
  entropy is even larger than before, panel (d), and they are compelled to
  continue drifting.
 The cycle
  continues, leading to instability. Notice that the cycle incorporates the $B$
  field, resistivity, and $dS/dR<0$, three ingredients which are represented
  in the stability discriminant in $a_0$, Eq.~\eqref{a0}.

The essential features of the instability are: the suppression of stabilising rotation by
the diffusing magnetic field, and slow (or negligible) thermal diffusion. This justifies
calling the instability double-diffusive: angular momentum is diffused
rapidly, eliminating the stable angular
momentum stratification. But heat is diffused slowly, allowing the unstable
entropy stratification to do its work unhindered.

 It can be shown that a sufficiently
large viscosity can do the same job as the magnetic field here: and, in fact, a
hydrodynamical disk in which $\xi/\nu\ll 1$ exhibits an analogous
instability (where $\nu$ is kinematic viscosity).
 Consequently, we may interpret the instability as the `dual'
 of the
Goldreich and Schubert instability (Goldreich and Schubert 1967), which relies
on the negation of a stabilising entropy gradient, by rapid thermal
diffusion, and the tapping of an unstable angular momentum gradient
(unimpeded by the weak viscosity). The dead-zone double-diffusion instability
is also related to salt-fingering in the ocean: and, in fact, the linear stage
of a single unstable mode will be characterised by a vertical sequence of
horizontal `fingers' or channel flows, much as in the MRI. However, these flows, unlike
classical MRI, are not nonlinear solutions, and will quickly break down into some
form of disordered mixing. Another important difference is that
nearly
 all the unstable modes
possess comparable growth rates, so no one mode will dominate. Noticeable channel flows,
or fingers, are therefore unlikely to be observed in simulations.

\section{Conclusion}

This paper introduces a novel linear instability which occurs on short length-scales
and which could be at work within the dead-zones of protostellar disks. The
instability is double-diffusive and relies on (a) the energy provided by a
weak negative entropy gradient, (b) a diffusing magnetic field to relax angular
momentum conservation, and (c) very slow thermal diffusion. It grows on a range of
lengthscales extending from the resistive length down to 
a length related to the thermal scale. Across this potentially large range the
growth
rate scales like
$s\sim\Lambda\,(N^2_R/\Omega^2)\,\Omega$. 
Though the unstable mode will almost always physically fit into the dead-zone
(unlike the MRI), it may grow quite slowly. Nevertheless it may give rise to dynamically
interesting velocity amplitudes after some $10^4$ orbits (depending on
the parameters).
In principle, the instability's nonlinear saturation may play a significant role in
 the dynamics of dead-zones, leading, in particular, to radial and vertical
 mixing, which in turn may influence dust grain dynamics (settling
 and clumping) on one
 hand, and the global temperature structure of the disk, on the other. 

 However, a number of issues need to
 be settled, and these form the basis for future work. First,
 the governing parameters of the system need to be better constrained, 
so that one can be
 assured the instability grows on dynamically
 meaningful timescales. This requires realistic estimates for the
 magnetic field strength, the radial entropy gradient, and the resistivity. 
  Second, the linear analysis should be extended to the peculiar nonideal MHD
 conditions present in the interior of protostellar disks, where the
 instability will be altered by Hall effects and ionisation and 
 dust grain chemistry. Third, the interaction between the instability and much
 faster processes in the active zones needs to be clarified. For instance, the turbulent
 motions in these regions may transport the large-scale magnetic
 field on a similar time to the growth time (Fromang and Stone 2009, Lesur and
 Longaretti 2009).
On the other hand, rapid density waves, stimulated by the turbulence, may also
 interact with the
 instability (Fleming and Stone 2003, Heinemann and Papaloizou 2009a,b).
 Fourth, the instability's nonlinear saturation should be simulated numerically
 in order to establish whether it leads to sustained disordered
 flows. 

Finally, this instability is not restricted
to protostellar disks, as the necessary ingredients for instability are 
rather generic. In particular, double-diffusive modes
 may be excited within the fluid cores of rapidly
rotating planets and protoplanets, possibly modulating pre-existing dynamo
activity, or perhaps directly instigating convection itself in bodies that
spin so rapidly that regular buoyancy instabilities are quenched. These issues
we address in future work.

\section*{Acknowledgements}

The authors would like to thank the anonymous referee
 whose critical suggestions
improved the quality of the manuscript.
HNL thanks Geoffroy Lesur, John
Papaloizou, and Pierre Lesaffre for much useful advice.
 Funding from the Conseil R\'egional de l'Ile
de France is acknowledged.

\appendix
\section{Asymptotic description of long wavelength modes}

In this appendix we calculate the growth rates of the five modes of
Eq.~\eqref{axisdispdim} in the limit of long wavelength, i.e.\ on scales longer than the
resistive scale $l_\eta$. 
As earlier, we
adopt $\Omega^{-1}$ for the unit of time, and the Alfv\'en wavenumber
($\Omega/v_A$) for the unit of wavenumber. 
In addition, we drop the thermal diffusion terms, assuming that
$k$ and $\xi$ are so small that these effects will be negligible.
 The dispersion
relation \eqref{axisdispdim} in these units is
\begin{equation} \label{axidisp}
s^5 + b_4\, s^4 + b_3\, s^3 + b_2\,s^2 + b_1\,s + b_0 =0,
\end{equation}
with coefficients
\begin{align*}
b_4 &= 2\Lambda^{-1}\,k^2, \\
b_3 &= \kappa^2+N^2_R + 2 k^2 +\Lambda^{-2}\,k^4,  \\
b_2 &= 2\,\Lambda^{-1}\,k^2\left(\kappa^2+N^2_R + k^2\right),    \\
b_1 &= k^2(k^2-4) +k^2(1+\Lambda^{-2}\,k^2)(N^2_R+\kappa^2), \\
b_0 &= \Lambda^{-1}\,N^2_R\,k^4,
\end{align*}
and in which both $\kappa^2$ and $N^2_R$ have been scaled by $\Omega^2$.

A small scaling parameter $\epsilon$ is introduced so that $0<\epsilon\ll 1$.
 From the outset we assume that both $\Lambda$ and $N^2_R$ are equally small,
 and let $\Lambda\sim N^2_R\sim  \epsilon$.
In order to pick out the longest wavelength modes we set $k\sim\epsilon^2$,
which means that we examine
lengthscales $\Lambda^{-1}$ times the resistive scale. Next the
growth rate $s$ is expanded in small $\epsilon$, which allows the computation
of each mode analytically to leading order

The two hydrodynamical convective modes possess an expansion like 
$s=s_0+s_1\,\epsilon + s_2\,\epsilon^2+\dots$, with the leading order term given
by
 $s_0= \pm i\,\kappa$.
Thus the modes oscillate at the epicyclic frequency at leading order. The
real part of the growth rate comes in only at \emph{seventh order} in $\epsilon$
\begin{equation}
 \text{Re}(s)= -\frac{4k^4}{\kappa^4\,\Lambda}+\mathcal{O}(\epsilon^8),
\end{equation}
and thus decays at a rate proportional to $\eta$.
Note that if a different scaling had
been adopted in which $N^2_R\sim 1$ then the leading order term would be
$ s_0= \pm \sqrt{-\kappa^2-N^2_R}$ and stability would follow the
Hoiland-Solberg criterion, $\kappa^2+N^2_R>0$.

The leading order term of the two MRI (or convective-MRI) modes scales as
$\epsilon^2$ and is determined by balancing the second and fourth term in
Eq.~\eqref{axidisp}. We find
\begin{equation}
s=\pm\frac{k}{\kappa}\sqrt{-\frac{d\Omega^2}{d\ln R}}\, +\,\mathcal{O}(\epsilon^4),
\end{equation}
which returns the MRI stability condition $d\Omega^2/d\ln R >0$.
 Thus these very long scales will be
MRI unstable. 
Note that the entropy gradient has not entered at leading order, it being too
small to determine the stability properties of the gas. But
if we had assumed that $N^2_R\sim 1$ the growth rate to leading order would be 
proportional to $\sqrt{-\kappa^2-N^2_R+4}$, and the instability criterion
would reproduce that for convective-MRI modes (Balbus 1995, see also Papaloizou and
Szuszkiewicz 1992).

The last mode is a simple resistive mode whose growth rate scales like
$\epsilon^4$ and which can be determined from the last two terms of the
dispersion relation. We find
\begin{equation} \label{resis}
s= -\frac{k^2\, N^2_R}{\Lambda\,(d\Omega^2/d\ln R)}\,+\,\mathcal{O}(\epsilon^6),
\end{equation}
and thus the mode decays at a rate proportional to $\eta$. Note that this mode
is \emph{not} the double-diffusive mode computed in the main body of the
paper. It exists on a separate branch of the dispersion relation and will
always decay for all $k$. However, just like the double-diffusive instability, it
establishes magnetostrophic balances in both the induction and momentum
equation (see Section 3.4.2). 

\end{document}